\documentclass[twocolumn,prl,amsmath,amssymb,showkeys,showpacs]{revtex4}
\usepackage{graphicx}
\newcommand{\be}{\begin{eqnarray}}
\newcommand{\ee}{\end{eqnarray}}

\newcommand{\ket}[1]{\vert{#1}\rangle}
\newcommand{\mtx}[3]{\langle{#1}\vert {#2}\vert{#3}\rangle}
\newcommand{\up}{\uparrow}
\newcommand{\down}{\downarrow}
\newcommand{\B}[1]{\mbox{\boldmath$#1$}}

\newcommand{\meq}{\mbox{=}}
\newcommand{\eqsn}[2]{
\hfill\break
\parbox{7.5cm}{
\begin{eqnarray*}
#2
\end{eqnarray*}
}
\hfill\parbox{0.3cm}{\begin{eqnarray}
\label{#1}
\end{eqnarray}}
}

\begin{document}
\title{Structure of the photon and magnetic field induced birefringence and dichroism}
\author{J.A.Beswick, C. Rizzo}
\affiliation{ Laboratoire Collisions, Agr\'egats, R\'eactivit\'e
(UMR 5589, CNRS - Universit\'e Paul Sabatier Toulouse 3), IRSAMC,
31062 Toulouse Cedex 9, France}
\date{\today}

\begin{abstract}
In this letter we show that the dichroism and ellipticity induced
on a linear polarized light beam by the presence of a magnetic
field in vacuum can be explained in the framework of the de
Broglie's fusion model of a photon. In this model it is assumed
that the usual photon is the spin 1 state of a
particle-antiparticle bound state of two spin 1/2 fermions. The other $S=0$ state is referred to as
the \emph{second} photon. On the other hand, since no charged
particle neither particles having an electric dipole are
considered, no effect is predicted in the presence of electric
fields and this model is not in contradiction with star cooling
data or solar axion search.
\end{abstract}

\maketitle

%\bigskip

%\section{Introduction}

Very recently an experimental observation of optical activity of
vacuum in the presence of a magnetic field has been reported
\cite{Zavattini:06} by the PVLAS collaboration. The observed
results could not be explained in the framework of the standard
Quantum ElectroDynamics (QED), and the existence of light
pseudoscalar spinless bosons of the same nature of the Peccei and
Quinn axion \cite{Peccei:77} has been suggested (see e.g.
\cite{Lamoreaux:06}). This explanation however is in contradiction
with other existing experimental data. In particular, the particle
needed to justify the PVLAS results should be largely produced in
the star core by interaction of photons with plasma electric
fields. Such a particle should escape because of its very low
coupling with matter, and induce a fast cooling of stars at a
level already excluded by astrophysical
observations\cite{Raffelt:96}. Moreover, CAST experiment
\cite{Zioutas:05} devoted to detect solar axions by conversion in
a magnetic field, has already excluded the existence of such a
particle in the range of mass and coupling constant necessary to
give the PVLAS effect. To get rid of this contradiction more
exotic solutions have been proposed. In particular, the existence
of a massive paraphoton which would couple with the standard
photon \cite{Masso:06} and with the axionlike particle (ALP), the
photon-initiated real or virtual production of pair of low mass
millicharged particles \cite{Gies:06}, and the existence of an
ultralight pseudo-scalar particle interacting with two photons and
a scalar boson and the existence of a low scale phase transition
in the theory \cite{Mohapatra:07}.\\

In this letter we show that dichroism and ellipticity induced on a
linear polarized light beam by the presence of a magnetic field in
vacuum can be predicted in the framework of the de Broglie's
fusion model of a photon \cite{Broglie:32}. In this model it is
assumed that the usual photon is the spin 1 state of a
particle-antiparticle bound state of two spin 1/2 fermions. The other $S=0$ state is referred to as
the \emph{second} photon. The mass of the usual photon is supposed
to be zero or negligible.

In particular, we show that taken the spin-spin coupling  and the
interaction with an external magnetic field  proportional to ${\B
s}_1.{\B s}_2$ and ${\B B}.({{\B s}_1}-{{\B s}_2})$,
respectively, magnetic induced birefringence and dichroism are
obtained with a    $(\B \epsilon.{\B B})^2$ pseudo-scalar
symmetry where ${\B \epsilon}$ is the polarization of the photon
(as usual ${\B \epsilon}$ is defined by the direction of the
electric field). Thus both dephasing and absorption appear for
linearly polarized light parallel to the external applied magnetic
field.

On the other hand, since no charged particle neither particles
having an electric dipole are considered, no effect is predicted
in the presence of electric fields and this model is not in
contradiction with star cooling data or solar axion search.\\

%\section{The model}

We consider the photon as composed of a spin 1/2 particle and its antiparticle. The spin Hamiltonian
is assumed to be approximate by
\begin{equation}\label{model_1}
H_0 = - \frac{\Delta}{\hbar^2}\,{\B s}_1.{\B s}_2
\end{equation}
with $\Delta>0$. The ground state eigenstates are then given by:
\begin{eqnarray}\label{model_2}
\ket{S\meq1, M_z\meq 1} &=& \ket{\up,\up}; \quad \ket{S\meq1, M_z\meq-1} = \ket{\down,\down}\nonumber
\\
\ket{S\meq1, M_z\meq0} &=& \frac{1}{\sqrt{2}}\,\Big(\ket{\up,\down}+\ket{\down,\up}\Big)
 \end{eqnarray}
with energy $E_1 = -\Delta/4$, corresponding to the ordinary
photon $\gamma_1$. The \emph{second} photon $\gamma_0$ is then
given by the excited singlet state
\begin{equation}\label{model_3}
\ket{S\meq0, M_z\meq0} = \frac{1}{\sqrt{2}}\,\Big(\ket{\up,\down}-\ket{\down,\up}\Big)
 \end{equation}
with energy $E_0 = (3/4)\,\Delta$. The energy difference between
the two photons $\gamma_1$ and $\gamma_0$ is then given by
$\Delta$.

We assume the particle/antiparticle have  magnetic moments $\B m_1
= (\beta\,\mu_B/\hbar)\,{\B s}_1$ and $\B m_2=-\B\mu_1$. Thus the total
magnetic moment ${\B m} = (\beta\,\mu_B/\hbar)\,({\B s}_1-{\B s}_2)$
has zero average value for the $\gamma_1$ photon and ${m}_z =
\beta\,\mu_B$ for the $\gamma_0$ photon.

In the presence of a magnetic field $\bf B$ along $Oz$ we shall have
\begin{equation}\label{model_4}
V = (\beta\,\mu_B\,B/\hbar)\,({{\B s}_1}_z-{{\B s}_2}_z)
 \end{equation}
 The only non-zero matrix element of $V$ is
\begin{equation}\label{model_5}
\mtx{S\meq1, M_z\meq0}{V}{S\meq0, M_z\meq0} = \beta\,\mu_B\,B
 \end{equation}

After diagonalisation
%\begin{eqnarray}\label{model_6}
\eqsn{model_6}{
%\overline{\ket{S\meq1, M_z\meq0}}
\ket{\Psi_{1,0}}
 &=& \mbox{cos}\theta\,\ket{S\meq1, M_z\meq0} + \mbox{sin}\theta\,\ket{S\meq0, M_z\meq0}
\\
%\overline{\ket{S\meq0, M_z\meq0}}
\ket{\Psi_{0,0}}&=& -\mbox{sin}\theta\,\ket{S\meq1, M_z\meq0} + \mbox{cos}\theta\,\ket{S\meq0, M_z\meq0}
}
%\end{eqnarray}

\noindent with
\begin{equation}\label{model_7}
\tan(2\,\theta) = 2 \beta\,\mu_B\,B/\Delta
 \end{equation}
and eigenvalues
%\begin{equation}\label{model_8}
\eqsn{model_8}{
\overline{E}_1 &=& \frac{E_1+E_0}{2} - \frac{\Delta}{2}\,\sqrt{1 + \tan^2(2\,\theta)}
\\
\overline{E}_0 &=& \frac{E_1+E_0}{2} + \frac{\Delta}{2}\,\sqrt{1 + \tan^2(2\,\theta)}
}
% \end{equation}
with the new energy difference
\begin{equation}\label{model_9}
\overline{\Delta} =  \Delta\,\sqrt{1 + \tan^2(2\,\theta)}
\end{equation}

%\section{Magnetic field action on a linearly polarized photon}

The ordinary photon $\gamma_1$ can be described by a linear
combination of the two helicity states $\ket{S\meq1, M_{{\B
k}}\meq\pm 1}$ where $M_{{\B k}}$ is the projection of the spin
angular momentum in the direction of propagation of the photon.
Let first note that if the $\gamma_1$ is propagating along the
direction of the magnetic field, only $\ket{S\meq1, M_z\meq\pm 1}$ will
be involved and no effect is expected. Consider now a $\gamma_1$
propagating along $Oy$ and linearly polarized in the direction of
$Oz$. We shall have
\begin{equation}\label{B_1}
\ket{\epsilon_z} = -\frac{1}{\sqrt{2}}\,\Big(
\ket{S\meq1, M_y\meq  1} - \ket{S\meq1, M_y\meq- 1}
\Big)
\end{equation}
but
\begin{equation}\label{B_2}
\ket{S, M_y} = \sum_{M_z \meq0, \pm1}\, \ket{S, M_z}\,d^S_{M_z,M_y}(\pi/2)
\end{equation}
where the $d^S_{M_z,M_y}$ are the Wigner $d$-functions.
Using
%\begin{equation}\label{B_3}
\eqsn{B_3}{
d^1_{1,\pm 1}(\Theta) &=& \frac{1}{2}(1\pm \cos\Theta)
\\ d^1_{0,\pm 1}(\Theta) &=& \pm\frac{1}{2}\,\sqrt{2}\,\sin\Theta
\\
 d^1_{-1,\pm 1}(\Theta) &=& \frac{1}{2}(1\mp\cos\Theta)
}
%\end{equation}

\noindent
we get from (\ref{B_1}) and (\ref{B_2})
\begin{equation}\label{B_4}
\ket{\epsilon_z} = - \ket{S\meq1, M_z\meq0}
\end{equation}
and this state will be affected by the magnetic field through its coupling to the $\ket{S\meq0,M_z\meq0}$ state.
We note in passing that in the case of a linear polarization along the $Ox$ axis we have
\begin{eqnarray}\label{B_5}
\ket{\epsilon_y} &=& \frac{1}{\sqrt{2}}\,\Big(
\ket{S\meq1, M_y\meq 1} + i\, \ket{S\meq1, M_y\meq- 1}
\Big)
\nonumber
\\&=& \frac{1}{\sqrt{2}}\,\Big(
\ket{S\meq1, M_z\meq  1} + i\,\ket{S\meq1, M_z\meq- 1}
\Big)
\end{eqnarray}
and this state will not be affected by the magnetic field.\\

%\subsection{Time dependence of the state vector}

We assume that the magnetic field is switch-on between $t=0$ and $t=\tau =L/c$, where
$L$ is the field length (of the order of 1 m in PVLAS experiment).
We shall have $\ket{\psi(0)} = -\ket{1,0} = - \cos\theta\,\ket{\Psi_{1,0}} + \sin\theta\,\ket{\Psi_{0,0}}$
where from now on the kets correspond to $\ket{S,M_z}$.
At time $\tau$
\begin{eqnarray}\label{time_1}
\ket{\psi(\tau)} &=&  -
\cos\theta\,e^{-i\,\overline{E}_1\,\tau/\hbar}\,\ket{\Psi_{1,0}}
\\
&+&
\sin\theta\,e^{-i\,\overline{E}_0\,\tau/\hbar}\,\ket{\Psi_{0,0}}
\end{eqnarray}
which  in terms of the non-perturbed kets $\ket{1,0}$ and $\ket{0,0}$, will be
given by
\begin{eqnarray}\label{time_2}
\ket{\psi(\tau)} &=&  - \left(\cos^2\theta\,e^{-i\,\overline{E}_1\,\tau/\hbar}+\sin^2\theta\,e^{-i\,\overline{E}_0\,\tau/\hbar}\right)\,{\ket{1,0}}
\nonumber\\
&-&
 \cos\theta\,\sin\theta\,\left(e^{-i\,\overline{E}_1\,t/\hbar} -e^{-i\,\overline{E}_0\,\tau/\hbar}\right)\,{\ket{0,0}}
\end{eqnarray}
This can be written as
%\begin{eqnarray}\label{time_3}
%\ket{\psi(\tau)} &=&  -e^{-i\,(E_1+E_0)\,\tau/2\hbar}\,\Bigg[\left( \cos^2\theta\,e^{i\,\overline{\Delta}\,\tau/2\hbar}
%+
%\sin^2\theta\,e^{-i\,\overline{\Delta}\,\tau/2\hbar}\right)\,{\ket{1,0}}
%\nonumber\\
%&+&
% \cos\theta\,\sin\theta\,\left(e^{i\,\overline{\Delta}\,\tau/2\hbar} -e^{-i\,\overline{\Delta}\,\tau/2\hbar}\right)\,{\ket{0,0}}\Bigg]
%\end{eqnarray}
%and also as
\begin{eqnarray}\label{time_4}
\ket{\psi(\tau)} &=&  -e^{-i\,(E_1+E_0)\,\tau/2\hbar}\,\Bigg[\Big( \cos(\overline{\Delta}\,\tau/2\hbar)
\nonumber
\\
&+&i\cos(2\theta)\,\sin(\overline{\Delta}\,\tau/2\hbar)\Big)\,{\ket{1,0}}
\nonumber\\
&+&
i \,\sin(2\theta)\,\sin(\overline{\Delta}\,\tau/2\hbar) \,{\ket{0,0}}\Bigg]
\end{eqnarray}

%\subsection{Magnetic induced dichroism}

From (\ref{time_4}) the probability to produce $\ket{0,0}$ is $P_{\gamma_1\to\gamma_0} = \vert\langle 0,0\ket{\psi(\tau)}\vert^2$
which gives
%$P_{\gamma_1\to\gamma_0} = \sin^2(2\theta)\,\sin^2(\overline{\Delta}\,\tau/2\hbar)$ and using
%(\ref{model_7}) and (\ref{model_9}),
\begin{equation}\label{dicro_1}
P_{\gamma_1\to\gamma_0} = \frac{ \tan^2(2\theta)}{1+\tan^2(2\theta)}\,\sin^2\Big({\Delta}\,\sqrt{1+\tan^2(2\theta)}\,\tau/2\hbar\Big)
\end{equation}
with $\tan^2(2\theta)$ given by (\ref{model_7}).
%\begin{equation}\label{dicro_2}
%\tan^2(2\theta) = \left(\frac{2 \beta\,\mu_B\,B}{\Delta}\right)^2
%\end{equation}

In the limit where $2 \beta\,\mu_B\,B\ll \Delta$, $\tan^2(2\theta)\ll 1$, and
%\begin{equation}\label{dicro_3}
%P_{\gamma_1\to\gamma_0} \simeq \left(\frac{2\,\beta\,\mu_B\,B}{\Delta}\right)^2\,\sin^2\Big(\Delta\,\tau/2\hbar\Big)
%\end{equation}
%If in addition $\Delta\,\tau/\hbar\ll \pi$,
%\begin{equation}\label{dicro_4}
%P_{(1,0)\to(0,0)} \simeq \left(\frac{ \beta\,\mu_B\,B\,\tau}{\hbar}\right)^2
%\end{equation}
%which can be also written as
\begin{equation}\label{dicro_3bis}
P_{\gamma_1\to\gamma_0} \simeq
\left(\frac{\beta\,\mu_B\,B\,\tau}{\hbar}\right)^2\,\left(\frac{\sin\Big(\Delta\,\tau/2\hbar\Big)}{\Delta\,\tau/2\hbar}\right)^2
\end{equation}
Thus, when $\Delta\,\tau/2\hbar \ll 1$, $P_{\gamma_1\to\gamma_0}$
does not depend on $\Delta$.

In an apparatus like the PVLAS one where a linearly polarized
laser passes through a region where a magnetic field pointing at
45 degrees with respect to light polarization plane is present,
such a conversion probability will show as a linear dichroism
giving an apparent rotation of the polarization plane $\rho =
\frac{1}{2} P_{\gamma_1\to\gamma_0}$. It is worth to stress that
standard QED \cite{Adler:71} does not predict any dichroism for
light propagating in vacuum in the presence of a magnetic field.
%\subsection{Magnetic induced birefringence}

As for the phase of the $\ket{1,0}$, this is given by
\begin{equation}\label{ellip_1}
\overline{\phi}_1 = -(E_1+E_0)\,\tau/2\hbar + \arctan\left[\cos(2\theta)\,\tan(\overline{\Delta}\,\tau/2\hbar)\right]
\end{equation}

On the other hand, for the $Ox$ polarization the phase is $\phi_1
= E_1\,\tau/\hbar$. The phase difference between $\gamma_1$ states
for polarization along and perpendicular to the magnetic field is
then given by
\begin{equation}\label{ellip_3}
\delta\phi \equiv \overline{\phi}_1 - {\phi}_1 = \arctan\left[\cos(2\theta)\,\tan(\overline{\Delta}\,\tau/2\hbar)\right] - {\Delta}\,\tau/2\hbar
\end{equation}

Expanding this function in powers of $\theta$ around zero, we found
\begin{equation}\label{ellip_4}
\delta\phi  = \left(\frac{\beta\,\mu_B\,B}{\Delta}\right)^2 \,\left(\frac{\Delta\,\tau}{\hbar}
- \sin({\Delta}\,\tau/\hbar)
\right)
\end{equation}

%If in addition $\Delta\,\tau/\hbar\ll \pi$, we have
%\begin{equation}\label{ellip_5}
%\delta\phi  = \frac{1}{6}\,\left(\frac{\beta \,\mu_B\,B}{\Delta}\right)^2\,\left(\frac{\Delta\,\tau}{\hbar}\right)^3=
%\frac{1}{6}\,\left(\beta \,\mu_B\,B\,\Delta\right)^2\,\left(\frac{\tau}{\hbar}\right)^3
%\end{equation}

%From (\ref{dicro_4}) and (\ref{ellip_5}), we note that we can write
%\begin{equation}\label{ellip_6}
%\delta\phi  =
%\frac{1}{6}\,P_{(1,0)\to(0,0)}\,\left(\frac{\Delta\,\tau}{\hbar}\right)
%\end{equation}

Again, in the case of an apparatus like the PVLAS one, this
dephasing will show as an ellipticity $\epsilon = \delta\phi/2$
acquired by the polarized beam passing through the magnetic field
region. Ellipticity is associated to the existence of a
birefringence by the formula

\begin{equation}\label{ellip_4bis}
\epsilon  = \frac{\pi\,L}{\lambda}(n_\|-n_\bot)
\end{equation}
where $\lambda$ is the light wavelength, and $n_\|$ and $n_\bot$
are the indexes of refraction for light polarized parallel and
perpendicular with respect to the magnetic field, respectively.
Thus, in the framework of our model a vacuum will show an apparent
magnetic birefringence

\begin{equation}\label{ellip_4ter}
(n_\|-n_\bot)=\left(\frac{\lambda}{2\pi\,c\,\tau}\right)
\left(\frac{\beta\,\mu_B\,B}{\Delta}\right)^2\,
\left(\frac{\Delta\,\tau}{\hbar}-\sin({\Delta}\,\tau/\hbar)\right)
\end{equation}
that depends on the time the photon stays in the magnetic field
region. Standard QED predicts that a vacuum is a magnetic
birefringent medium showing a $(n_\|-n_\bot) \simeq 4\times
10^{-24} B^2$ where $B$ is given in Tesla. That only depends on
the value of fundamental constants and the square of the magnetic
field intensity \cite{Adler:71}. QED also predicts that a
corresponding effect exists in the presence of an electric field,
such an effect is absent in the framework of our model.\\

%\section{Discussion}

We note that our formulas for the conversion probability and the
dephasing are equivalent to the ones obtained in the axion case
\cite{Maiani:86} since axion-photon coupling can be also treated
as a two level system \cite{Raffelt:88}. Our $\Delta$ corresponds
to the ratio ${m_a^2}/{\omega}$ and $\beta$ to
$g_{a\gamma\gamma}$, where $m_a$ is the axion mass, $\omega$ the
photon energy, and $g_{a\gamma\gamma}$ the axion-photon coupling
constant. The mass and coupling constant associated to the ALP
needed to explain PVLAS results, $m_a \approx 10^{-3}$ eV and
$g_{a\gamma\gamma} \approx 3 \times 10^{-6}$ GeV$^{-1}$
\cite{Zavattini:06}, have been chosen by comparing the dichroism
signal of PVLAS with limits published by the BRFT collaboration in
1993 \cite{Cameron:93}. In fig. 1 we show the corresponding graph
following equation (\ref{dicro_3bis}). We have assumed as usual that
the measured effect is simply the effect predicted by this formula
multiplied by the number of passages in the magnetic field due of
the presence of optical cavities. Dotted line represents the lower
border of the parameters plane forbidden by BRFT results at a
2$\sigma$ level, while full line represents the PVLAS signal.

The main difference between our model and the axion model is that
in our case the optical effects do not depend on the photon
energy. Thus, in our case, the oscillations between the two states
of the hamiltonian only depend on the time the $\gamma_1$ stay in
the magnetic field i.e. the length of the magnetic field region.
In the axion case the oscillations depend on the length divided by
the photon energy $\omega$. Oscillations can therefore be avoided
by choosing higher energy photons for longer magnets, and that was
the case of BRFT collaboration with respect to PVLAS
collaboration. Eventually, this explains why in our case the
allowed window for the parameters that could explain the PVLAS
dichroism is larger that in the axion case treated in ref.
\cite{Zavattini:06}.\\

\begin{figure}
  \includegraphics[width=9cm]{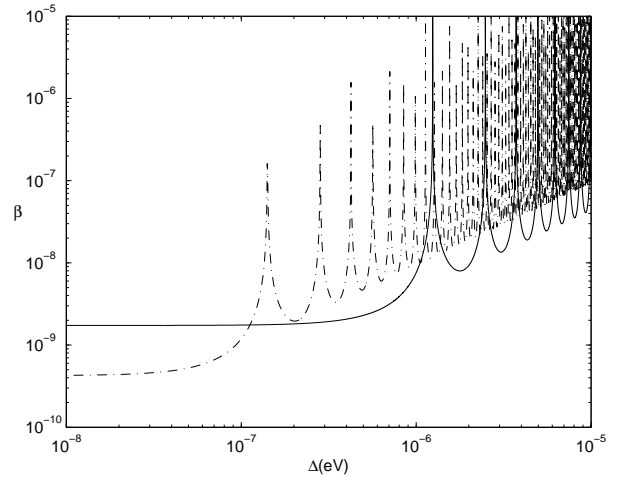}\\
  \caption{Comparison between PVLAS signal and BRFT limits}\label{fig1}
\end{figure}

In our model the mixing between the ordinary photon $\gamma_1$ and
the second photon $\gamma_0$ only appears in a magnetic field.
This will not affect the energy balance and star evolution, but
should be important in the case of photon emission from neutron
stars which show magnetic fields as high as $10^9$ T. This is
anyway an important issue also for ALP (see e.g. ref.
\cite{Dupays:05}, and \cite{Lai:06}).\\

Since the fusion model assumes  structure of the photon, we should
expect excited states associated to internal motion and presumably
non zero mass of the constituents. However, if the masses and the
spatial dimension of the photon are very small, the first excited
state can be very high up in energy.\\

In conclusion, once the PVLAS signal will be confirmed, the exotic
but simple de Broglie's fusion model for the photon can provide an
explanation for this signal that is not in contradiction with star
observation or solar axion search. On the other hand, experiments
testing the propagation of light in the presence of a magnetic
field in terrestrial laboratories or by astrophysical observations
can put more and more stringent limits on its free parameters.

\section{Acknowledgements}

We thanks E. Mass\'o for very helpful discussion and carefully
reading our manuscript.
%
%Bibliography%
%%%%%%%


\begin{thebibliography}{10}

\bibitem{Zavattini:06} E. Zavattini {\it et al.}, {\it Phys. Rev. Lett.} {\bf 96},
110406 (2006).

\bibitem{Peccei:77} R. D. Peccei and H. R. Quinn, {\it Phys. Rev. Lett.} {\bf 38},
1440 (1997).

\bibitem{Lamoreaux:06} S. Lamoreaux, {\it Nature} {\bf 441},
31 (2006).

\bibitem{Raffelt:96} G. G. Raffelt, {\it Stars as Laboratories for
Fundamental Physics} (University of Chicago Press, Chicago, 1996).


\bibitem{Zioutas:05} K. Zioutas {\it et al.}, {\it Phys. Rev. Lett.} {\bf 94},
121301 (2005).

\bibitem{Masso:06} E. Masso and J. Redondo, {\it Phys. Rev. Lett.} {\bf
97}, 151802 (2006).

\bibitem{Gies:06} H. Gies, J. Jaeckel and A. Ringwald, {\it Phys. Rev. Lett.} {\bf 97},
140402 (2006).


\bibitem{Mohapatra:07} R. N. Mohapatra and Salah Nasri, {\it Phys. Rev.
Lett.} {\bf 98}, 050402 (2007).

\bibitem{Broglie:32}
L. de Broglie, C. R. Acad. Sci. Paris {\bf 195}, 536 (1932);  {\bf
195}, 577 (1932); {\bf 197}, 1377 (1933); {\bf 198}, 135 (1934).

\bibitem{Adler:71} S. L. Adler, {\it Ann. Phys. (NY)} {\bf 67}, 599
(1971).

\bibitem{Maiani:86} L. Maiani, R. Petronzio and E. Zavattini, {\it Phys. Lett. B} {\bf 175},
359 (1986).

\bibitem{Raffelt:88} G. Raffelt and L. Stodosky, {\it Phys. Rev. D} {\bf 37},
1237 (1988).

\bibitem{Cameron:93} R. Cameron {\it et al.}, {\it Phys. Rev. D} {\bf 47},
3707 (1993).

\bibitem{Dupays:05} A. Dupays, C. Rizzo, M. Rocandelli, and G.F. Bignani {\it Phys. Rev. Lett.} {\bf 95},
211302 (2005).

\bibitem{Lai:06} D. Lai and J Heyl, {\it Phys. Rev. D} {\bf 74},
123003 (2006).

\end{thebibliography}
\end{document}